\documentclass[aps,prl,twocolumn,showpacs,groupedaddress,nofootinbib,preprintnumbers]{revtex4}

\usepackage{epsfig}
\usepackage{amsmath}
\usepackage{axodraw}
\usepackage{color}

\newcommand{\beq}{\begin{equation}}
\newcommand{\bea}{\begin{eqnarray}}
\newcommand{\eeq}{\end{equation}}
\newcommand{\eea}{\end{eqnarray}}
\newcommand{\eR}{e_{\text{R}}}
\newcommand{\eL}{{e}_{\LUV}}
\newcommand{\eLq}{{e}_{\LUV}^2}

\newcommand{\eRq}{e_{\text{R}}^2}
\newcommand{\mR}{m_{\text{R}}}
\newcommand{\mL}{{m}_{\LUV}}
\newcommand{\LUV}{\Lambda}
\newcommand{\LL}{\Lambda_{\text{L}}}
\newcommand{\xsb}{$\chi$SB}
\newcommand{\Nf}{N_{\text{f}}}
\newcommand{\Gk}{\Gamma_k}
\newcommand{\Gt}{\Gamma_k^{(2)}}
\newcommand{\pat}{\partial_t}
\newcommand{\STr}{\,\text{STr}\,}
\newcommand{\yb}{\bar{\psi}}
\newcommand{\ZF}{Z_F}
\newcommand{\Zy}{Z_\psi}
\newcommand{\Zp}{Z_\phi}
\newcommand{\eF}{\eta_F}
\newcommand{\ey}{\eta_\psi}
\newcommand{\rF}{r_{\text{F}}}
\newcommand{\I}{\text{i}}
\newcommand{\fss}[1]{#1\!\!\!/}
\newcommand{\fsl}[1]{#1\!\!\!\!/}
\newcommand{\re}[1]{~(\ref{#1})}
\newcommand{\blp}{\bar{\lambda}_{+}}
\newcommand{\blm}{\bar{\lambda}_{-}}
\newcommand{\VAp}{\,(\text{V+A})}
\newcommand{\VAm}{\,(\text{V--A})}
\newcommand{\Li}{\text{L}\!\text{i}}
\newcommand{\eqcr}{e^2_{\text{cr}}}

\begin{document}

\preprint{HD-THEP-04-21}

\title{Renormalization flow of QED}

\author{Holger Gies}
\author{Joerg Jaeckel}
\affiliation{Institute for Theoretical Physics, Heidelberg University\\
Philosophenweg 16, 69120 Heidelberg, Germany}

\begin{abstract}
\noindent
We investigate textbook QED in the framework of the exact
renormalization group. In the strong-coupling region, we study the
influence of fluctuation-induced photonic and fermionic
self-interactions on the nonperturbative running of the gauge
coupling. Our findings confirm the triviality hypothesis of complete
charge screening if the ultraviolet cutoff is sent to infinity. Though
the Landau pole does not belong to the physical coupling domain owing
to spontaneous chiral symmetry breaking (\xsb), the theory predicts a
scale of maximal UV extension of the same order as the
Landau pole scale. In addition, we verify that the \xsb\ phase of the
theory which is characterized by a light fermion and a Goldstone boson
also has a trivial Yukawa coupling.
\end{abstract}
\pacs{12.20.-m, 11.15.Tk, 11.10.Hi}

\maketitle

Though quantum field theory celebrates its greatest triumph with
quantum electrodynamics (QED), the high-energy behavior of QED remains
a sore spot, since it is inaccessible to the otherwise successful
perturbative concepts. For instance, keeping the renormalized coupling
$\eR$ fixed, small-coupling perturbation theory predicts its own
failure in the ultraviolet (UV) in the form of the Landau pole singularity:
\begin{equation}
\frac{1}{\eRq}-\frac{1}{\eLq}=\beta_0\, \ln
\frac{\LUV}{\mR}, \quad \beta_0=\frac{\Nf}{6\pi^2}. \label{Lpole}
\end{equation}
The coupling $\eL$ at the UV cutoff $\LUV$ diverges for
$\LUV\to\LL=\mR \exp(1/(\beta_0 \eRq))$. It was early realized
\cite{Landau} that this behavior can signal the failure of QED as a
fundamental quantum field theory which should be valid on all length
scales. From a different viewpoint, keeping the initial UV coupling $\eL$
fixed, the renormalized coupling $\eR$ vanishes in the limit $\LUV\to
\infty$, resulting in a free, or ``trivial'', theory with complete
charge screening.

Already in the dawning of the renormalization group (RG), a possible
alternative scenario was discussed \cite{Gell-Mann:fq} in which
an interacting UV-stable fixed point of the RG transformation,
$\eLq\to e_\ast^2\in (0,\infty)$ for $\LUV\to\infty$, facilitates a
finite UV completion of QED (``asymptotic safety''
\cite{Weinberg:1976xy}). However, no sign of such a fixed point has
been found so far. On the contrary, nonperturbative lattice
simulations have provided evidence for triviality
\cite{Gockeler:1997dn,Kim:2000rr}. Moreover, careful extrapolation of
raw lattice data shows that the Landau pole singularity is outside the
physical parameter space owing to the onset of spontaneous chiral
symmetry breaking (\xsb) \cite{Gockeler:1997dn}. This strong-coupling
phenomenon of \xsb\ has also been observed in analytical studies using
truncated Dyson-Schwinger equations (DSE) in a quenched approximation
\cite{Miransky:1984ef}.

In addition to the fundamental character of this problem as a matter
of principle, the high-energy fate of QED or other standard-model
building blocks and its further extensions can give us direct bounds
on the scale where new physical phenomena may be expected. In
particular Landau pole singularities of the type of Eq.\re{Lpole} are
used to constrain properties of hypothetical particles, such as the
Higgs scalar in the standard model \cite{Hambye:1996wb}. Our work is
moreover motivated by the recent observation that a hypothetically
nontrivial U(1) sector of the standard model with a UV-stable fixed
point has the potential to solve the hierarchy problem of the Higgs
sector \cite{Gies:2003dp}.

In this letter, we report on nonperturbative results obtained from the
RG flow equation for the effective average action $\Gk$
\cite{Wetterich:1993yh}. We work in Euclidean spacetime continuum where our
methods can easily bridge a wide range of scales, allow for the full
implementation of chiral symmetry as well as a simple inclusion of
bare masses (explicit \xsb\ terms), and furnish unquenched
calculations.

The effective average action is a free-energy functional that
interpolates between the initial UV action $\Gamma_{k=\LUV}$ and
the full quantum action $\Gamma_{k\to 0}$. The infrared (IR) regulator
scale $k$ separates the fluctuations with momenta $p^2\gtrsim k^2$,
the effect of which has already been included in $\Gk$, from those with
smaller momenta which have not yet been integrated out. The full RG
trajectory is given by the solution to the flow equation ($t=\ln (k/\LUV)$),
\begin{equation}
\pat\Gk[\phi]=\frac{1}{2} \STr \pat R_k\, (\Gt[\phi]+R_k)^{-1},
\label{floweq}
\end{equation}
where $\Gt$ denotes the second functional derivative with respect to
the fields $\phi=(A_\mu,\yb,\psi)$,
and the regulator function
$R_k$ implements the infrared regularization at $p^2\simeq
k^2$. Effectively, Eq.\re{floweq} is a smooth realization of the
Wilsonian momentum-shell integration, being dominated by momenta
$p^2\simeq k^2$.

On microscopic scales, QED is defined by the action
\begin{equation}
S_\Lambda=\int_x \left( \frac{1}{4} F_{\mu\nu}F_{\mu\nu} +\yb
\I\fsl{D}[A] \psi +\yb \gamma_5 \mL \psi\right), \label{SLambda}
\end{equation}
which involves the microscopic UV parameters $\eL$ and $\mL$, and
$D_\mu[A]=\partial_\mu -\I \eL A_\mu$. Further possible
gauge - invariant interactions are RG irrelevant by
power counting. Invoking the universality hypothesis, the IR physics
should only depend on the parameters occurring in Eq.\re{SLambda}.
This hypothesis can nevertheless be questioned:
 since the coupling
increases towards the UV, higher operators can acquire large anomalous
dimensions that spoil naive power counting and enlarge the set of RG
relevant operators, offering new routes to UV completion. These
operators may in turn exert a strong influence on the running gauge
coupling and potentially induce an interacting fixed point. In order
to test this scenario quantitatively, we study how the running of the
gauge coupling can be affected by photonic self-interactions of the
type
\begin{eqnarray}
\!\!\Gamma_{k,A}\!\!&=&\!\!\!\int_x\! W(\theta)
=\!\!\int_x\!\left( W_1 \theta+\frac{W_2}{2}
   \theta^2+\frac{W_3}{3!}\theta^3+ \dots\right)\!\!, \label{truncA}
\end{eqnarray}
where $\theta=({1}/{4}) F_{\mu\nu} F_{\mu\nu}$.  Thus, we include
infinitely many fluctuation-induced photon operators in our truncation
of $\Gk$ ($W_1\equiv\ZF$ denotes the wave function renormalization of
the photon). Of course, there are further tensor structures involving,
e.g., the dual field strength that can contribute to the flow, but we
do not expect their influence on the running coupling to be
qualitatively different from those of Eq.\re{truncA}. Moreover, our
truncation neglects the momentum dependence of the couplings
$W_i$. Since it is natural to assume that their strength will
drop off with increasing external momenta, we expect that momentum
dependencies imply a weaker influence on the gauge coupling than is
estimated by Eq.\re{truncA}. Note that this argument could be
invalidated by the occurrence of yet unknown photonic bound states
giving rise to momentum poles in the couplings $W_i$.

Fluctuations induce not only photonic but also fermionic
self-interactions, the lowest order of which we include in the
fermionic part of the truncation,
\begin{eqnarray}
\Gamma_{k,\psi}&=& \int_x\bigg(
\yb(\I\Zy\fss{\partial}+Z_{1}\eL\fsl{A}+ \Zy m \gamma_5 )\psi\label{truncpsi}\\
&&   +\frac{1}{2} \big[ Z_{-}\blm \!\VAm + Z_{+}\blp \!\VAp \big]\bigg)
\nonumber,
\end{eqnarray}
where $(\text{V}\!\pm\!\text{A}):=\!(\yb\gamma_\mu\psi)^2 \mp
(\yb\gamma_\mu\gamma_5\psi)^2$. These fermionic interactions do
not only influence the running of the gauge coupling but are also
essential for the approach to \xsb\ in reminiscence of the
Nambu--Jona-Lasinio (NJL) model. The $k$-dependent dimensionless running
couplings are related to the bare couplings $\eL,\bar{\lambda}_{\pm}$ by
\begin{equation}
e=\frac{\eL Z_{1}}{\ZF^{{1}/{2}}Z_{\psi}},
\quad {\lambda}_\pm=\frac{Z_{\pm}k^{2}\bar{\lambda}_\pm}{\Zy^2}.
\label{runningc}
\end{equation}
QED initial conditions for the flow are defined by
%
$\ZF,\Zy,Z_1\big|_{\LUV}=1$ and 
$W_{i>1},\bar\lambda_\pm\big|_{\LUV}=0$. 
%

Inserting this truncation into Eq.\re{floweq}, we obtain the $\beta$
functions for $e$, ${\lambda}_\pm$, $m$, $\ZF$, $W_i$ and $\Zy$, once
the regulator $R_k$ is specified. Of central interest is the photon
anomalous dimension $\eF=-\pat\ln\ZF$ which contains the photon
self-interaction contributions to the $\beta_{e^2}$ function,
$\beta_{e^2}=\eF e^2+\dots$ (cf. Eq.\re{beta} below).
In order to deal with the photon sector of Eq.\re{truncA},
we use techniques developed in \cite{Gies:2002af} that employ
background-field-dependent and chiral-symmetry-preserving regulators
of the form
\begin{eqnarray}
R_{k}^{\psi}(\I \fsl{\bar D})&=&\Zy \I \fsl{\bar D}\,\,\,\,\,
    \rF\big((\I \fsl{\bar D})^2/k^2\big), \nonumber\\
R_{k}^{A}(\bar \Gamma^{(2)}_{k,A})&=& \bar\Gamma^{(2)}_{k,A}\,\,\,\,
    r\big(\bar\Gamma^{(2)}_{k,A}/(\ZF k^2)\big),
 \label{regulator}
\end{eqnarray}
where the bar indicates background-field dependence, and $r(y),\rF(y)$
denote dimensionless regulator shape functions. As a result, we arrive
at an asymptotic series for $\eF$ to all orders of the coupling,
\begin{displaymath}
\eF=\sum_{n=1}^\infty a_n(r;m,\ey)\,
\left(\frac{e^2}{16\pi^2}\right)^n =\frac{\Nf}{6\pi^2}
e^2+\mathcal{O}(e^4, m^2e^2), 
\end{displaymath}
where the coefficients $a_n$ depend functionally on the regulator
shape functions $r,\rF$. Here, the structure of the all-order result
arises from the feedback of the flow of the $W_i$'s on $\eF$, whereas
the global shape of the function $W(\theta)$ has been neglected
\cite{Gies:2002af}. To one loop, we obtain the correct universal
$\beta_{e^2}$ function coefficient, since $\beta_{e^2}=\eF
e^2+\dots$. To higher order, the result is explicitly regulator
dependent as it should be, since only the existence of zeroes of the
$\beta_{e^2}$ function and their critical exponents are
universal.\footnote{Already the two-loop coefficient is regulator
dependent, since we are using a mass-dependent regularization scheme.}
Now, QED could evade triviality if a UV-stable fixed point in
$\beta_{e^2}$ and $\eF$ existed for all regulators. By contrast, our
results show that $\eF(e^2_\ast)=0$ has only the solution $e_\ast^2=0$
for {\em all} regulator shape functions $r,\rF$.
In fact, for all physically admissible regulators a lower bound
$0<{\eF^{\textrm{1-loop}}}/{2}\leq\eF[r]$ exists for all values $e^2>0$.
In the strong-coupling regime, this lower bound is
satisfied by Litim's optimized regulator \cite{Litim:2001up},
\begin{displaymath}
\rF(y)=\frac{1}{\sqrt{y}}(1-\sqrt{y})\theta(1-y),\quad
r(y)=\frac{1}{{y}}(1-{y})\theta(1-y),\label{litim}
\end{displaymath}
for which the all-order anomalous dimension yields a simple integral
representation,
\begin{eqnarray}
\eF&=& \frac{e^2 \Nf}{6\pi^2} \, \frac{1-\ey}{1+m^2/k^2} \big[1 -
    I(e^2)], \label{etaF2}\\
&& I(e^2)=\frac{1}{\pi^2}\int_0^\infty ds\,
    s^2\, K_2(2\sqrt{s})\,\Li_2\left( e^{-\frac{4\pi^2}{e}
    \sqrt{\frac{3}{s}}}\right), \nonumber
\end{eqnarray}
involving a modified Bessel function $K_2$ and the polylogarithm
$\Li$. In the strong-coupling limit, $e^2\to\infty$, the integral goes
to $I(e^2)\to 1/2$, such that the strong-coupling limit finally
approaches half the one-loop result.\footnote{The fact that the
strong-coupling result can be expressed in terms of the one-loop
result only is likely to be accidental for the optimized
regulator. For instance, the frequently used exponential regulator
implies a strong-coupling behavior of the form $\eF\sim e^3$ which
cannot be expressed in terms of perturbative contributions only.}
Moreover, the explicit electron mass dependence illustrates the
threshold behavior: once the IR scale $k$ drops below the electron
mass scale, fluctuations become strongly suppressed and the flow
essentially stops.

The fermionic self-interactions also contribute directly to the
$\beta_{e^2}$ function. The detailed form can be read off from a
Ward-Takahashi identity as demonstrated in \cite{Gies:2003dp},
%
\begin{equation}
\pat e^2\equiv\beta_{e^2}= \eF e^2 +2 e^2\frac{\sum_{i=\pm} c_i \pat
\lambda_i}{1+\sum_{i=\pm} c_i \lambda_i},
\label{beta}
\end{equation}
where $c_+=\Nf/(4\pi)^2$, $c_-=(\Nf+1)/(4\pi)^2$ for the optimized
regulator. From this representation, it is apparent that if $\eF$ does
not induce a UV-stable interacting fixed point, no such fixed point
can be induced at all, since $\pat\lambda_\pm\!\!\to\! 0$ at a global
fixed point. (Explicit representa\-tions of the $\lambda_\pm$ flows can
be found in \cite{Gies:2003dp}.) As one of our main results, we
therefore exclude such a fixed point for the resolution of the
triviality problem. We have confirmed that even higher-order fermionic
and fermion-gauge field interactions cannot modify the qualitative
structure of Eq.\re{beta}.  The gauge coupling hence is generally not
bounded from above for increasing UV cutoff.

In order to deal with the phenomenon of \xsb\ that we expect for
strong-coupling, we use partial bosonization techniques as developed
in \cite{Gies:2001nw} in order to study the formation of the chiral
condensate and a dynamical fermion mass. Moreover, we can treat
dynamical as well as explicit fermion masses on the same footing by
translating the fermion self-interactions as well as the fermion mass
into a bosonic sector of the form
\begin{equation}
\Gamma_{k,\phi}=\int \Zp|\partial_\mu \phi|^2 +U(\phi)
+ \bar h (\bar{\psi}_{\text{R}} \psi_{\text{L}} \phi
-\bar{\psi}_{\text{L}} \psi_{\text{R}} \phi^\ast  ). \label{truncBos}
\end{equation}
Here we concentrate on the scalar boson in the $s$ channel. We
truncate the scalar potential to the simple form
\begin{equation}
U(\phi)=\bar m_\phi^2 \phi^\ast\phi +\frac{1}{2} \bar \lambda_\phi
(\phi^\ast \phi)^2 -\frac{1}{2}\bar \nu
(\phi+\phi^\ast),\label{Uofphi}
\end{equation}
where the $\bar\nu$ term breaks chiral symmetry explicitly and thus
carries the information about an explicit electron mass; if
$\bar\nu=0$ vanishes at any scale it vanishes at all scales by chiral
symmetry (massless QED). Spontaneous \xsb\ is monitored by the sign of
$\bar m_\phi^2$, negative values indicating an induced chiral
condensate.

Following the techniques of \cite{Gies:2001nw}, we trade the
four-fermion interactions and the electron mass for the parameters
occurring in Eq.\re{truncBos}, such that, for instance, the resulting
electron mass can be deduced from
\begin{equation}
m=\bar{h}|\phi_0| / \Zy, \label{emass}
\end{equation}
where $\phi_0$ denotes the minimum of the potential\re{Uofphi}. We
would like to stress that the fermion-boson translation employed here
is a highly efficient technique for controlling the (approximate)
chiral symmetry together with its explicit breaking by the mass; no
fine-tuning of the bare mass is necessary and there is no
proliferation of symmetry-breaking operators as in a purely fermionic
formulation. Together with the $\beta$ functions for the bosonic
sector (see \cite{Gies:2002hq}), we can evaluate the RG trajectory of
the complete system for a variety of initial conditions. Although the
number of parameters has seemingly increased, the system remains
solely determined by the choice of the gauge coupling and the electron
mass,
owing to the existence of an IR stable
``bound-state'' fixed point \cite{Gies:2001nw,Gies:2002hq}. This is a
manifestation of universality: the physics at large distance scales is
independent of the details of the microscopic interactions.

For the quantitative analysis, we work in the Landau gauge which is a
fixed point of the RG, and we concentrate on the $\Nf=1$ case where
the ``chiral'' symmetry is given by $U_{\text{F}}(1)\times
U_{\text{A}}(1)$, i.e., fermion-number and axial $U(1)$'s with \xsb\
corresponding to the breaking of $U_{\text{A}}(1)$.  At zero bare
mass, $\mL=0$, i.e., without explicit \xsb, our analysis reveals two
phases separated by a critical coupling $\eqcr$. For $\eLq\leq\eqcr$,
chiral symmetry is preserved and the electron remains massless. For
$\eLq>\eqcr$, \xsb\ renders the electron massive and a Goldstone boson
arises from the $\phi$ field. Switching on an explicit electron mass,
the transition between the two phases turns into a crossover with the
light mode of the $\phi$ field interpolating between a positronium
bound state and a pseudo-Goldstone boson.

In our truncation, the value of the critical coupling is
$\eqcr=38.41$. For comparison, we also mention the result for $\eqcr$
in the quenched approximation, $\eqcr{}_{,\text{q}}\simeq 14.81$,
which is in reasonable agreement with the quenched DSE result
\cite{Miransky:1984ef} in the Landau gauge,
$\eqcr{}_{,\text{qDSE}}=4\pi^2/3\simeq13.16$.  Note that our
approximation includes non-ladder diagrams such that gauge-dependences
are reduced \cite{Aoki:1996fh}.
The relation
$\eqcr>\eqcr{}_{,\text{q}}$ results from the fact that unquenched
fluctuations imply charge screening; therefore larger bare couplings
are necessary for \xsb.

\begin{figure}[t]
\begin{center}
\scalebox{0.8}[0.8]{
\begin{picture}(350,185)
\Text(00,165)[c]{\scalebox{1.5}[1.5]{$\frac{1}{e^{2}}$}}
\Text(260,-10)[c]{\scalebox{1.2}[1.2]{$\log_{10}(m/\LUV)$}}
\includegraphics{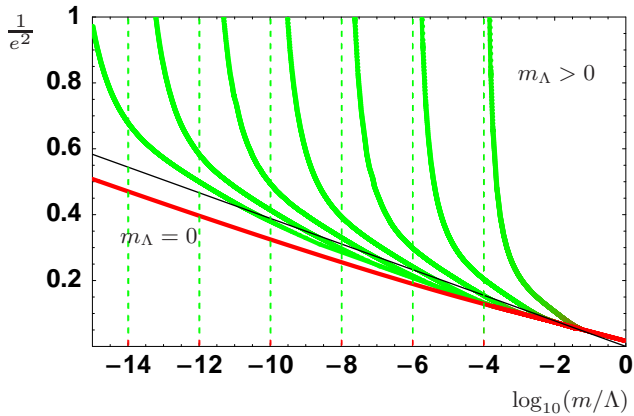}
\Text(-230,68)[c]{\scalebox{1.2}[1.2]{$\mL=0$}}
\Text(-42,145)[c]{\scalebox{1.2}[1.2]{$\mL>0$}}
\end{picture}}
\end{center}
\caption{Map of the 
bare couplings $({1}/{\eLq}, \log_{10}(\mL/\LUV))$ to the plane of
renormalized couplings
$({1}/{e^{2}_{\textrm{R}}},\log_{10}(m_{\textrm{R}}/\LUV))$. The
dashed vertical lines denote lines of constant bare mass in the
bare coupling plane which are mapped onto the solid lines in the
renormalized coupling plane (sub- and supercritical values of the
bare coupling are denoted by green and red, respectively).
The solid red line is the line of vanishing bare
mass (the thin black line, its 1-loop counterpart). Its pre-image
is a vertical line at $-\infty$. Note that the region below this
line is inaccessible, i.e., for a certain fixed value of the
renormalized coupling we have a minimal renormalized mass in units
of the cutoff. Hence it is impossible to send the cutoff to
infinity while keeping both renormalized mass and coupling fixed.
This demonstrates triviality.} \label{fig1}
\end{figure}

In Fig.\ref{fig1}, we plot the resulting renormalized values of the
gauge coupling and electron mass,
\begin{equation}
\eR=\lim_{k\to0} e, \quad \mR=\lim_{k\to 0}m, \label{rencoup}
\end{equation}
as functions of the bare parameters. Shown are lines of constant
bare mass $\mL$. For finite $\mL$, the curves exhibit a
linear regime and a pole. This displays the crossover behavior from a
\xsb\ dominated mass at strong coupling (linear regime) to an explicit
mass term at weak coupling; the limiting pole corresponds to
$\mR\simeq\mL$ for weak coupling. If we attempt to move the cutoff to
infinity but keep $\mR$ fixed, we need to take the limit $\mR/\LUV\to
0$ which can only be approached on the curve $\mL/\LUV\to 0$. In this
limit $\ln (\mR/\LUV)\to -\infty$, the renormalized coupling
goes to $\eR\to0$. This is the manifestation of triviality: the
whole range of bare couplings $0\leq\eLq\leq\eqcr$ for $\mR$ fixed
is mapped onto a single point $\eRq=0$. For a nontrivial theory, at
least one curve would have to intersect the $1/\eRq$ axis at some
finite $\eRq$ for $\mR/\LUV\to 0$.

On the other hand, if we want to keep $\eR>0$ fixed, we are forced
to accept a finite value for $\mR/\LUV>0$. Fixing the electron
mass to its physical value also determines the absolute value of
the cutoff, once the bare mass is fixed. The maximal cutoff value
is obtained for vanishing bare mass $\mL/\LUV\to0$, and we find
$\Lambda_{\text{max}}^{\mL=0}\sim 10^{278\pm 8}$GeV for QED
parameters. Yet, the limit $\mL\to0$ does not correspond to
``ordinary'' QED, since the electron mass is then fully generated
by \xsb, and a massless Goldstone boson arises. In order to
rediscover ``ordinary'' QED in the IR with given $\eR$ and $\mR$,
we have to choose a sufficiently large bare mass $\mL$ in order to
lift the pseudo-Goldstone boson to a positronium state with
mass $\simeq 2\mR$. This implies a small reduction of the maximal
UV scale.

For given renormalized mass and coupling, we observe that the
maximum possible bare coupling $\eL$ occurs for $\mL\to 0$, which
is a supercritical but still finite number. This fact describes
the absence of the Landau pole singularity: for given physical IR
parameters, large bare coupling values are inaccessible owing to
\xsb, in agreement with \cite{Gockeler:1997dn}.

We would like to stress that the maximal UV scale is regulator dependent. Considering
QED as being embedded in an underlying theory, the latter should become
visible at this scale. In this sense, the regulator dependence corresponds to the
physical threshold behavior towards the underlying theory.

Next we check whether QED can evade triviality in an unusual way:
we fine-tune the system onto $\eqcr$ from above with
$\mR/\Lambda\to 0$, such that the IR
spectrum consists of a light fermion, a free photon (since $\eR\to
0$), and a Goldstone boson with Yukawa coupling to the fermion. In
other words, QED with \xsb\ could have a Yukawa theory as
low-energy limit. However, we have confirmed explicitly that this
Yukawa coupling is also trivial in the limit of $\LUV\to \infty$
in much the same way
as the gauge coupling.

We would finally like to point to open questions of the present
investigation. First, our truncation of the fermion sector is
organized as a derivative expansion. This is justified if the fermion
anomalous dimension $\ey$ remains small. In the Landau gauge, we have
confirmed that this is indeed the case even at strong coupling, so our
truncation is self-consistent. Nevertheless, as is visible in
Eq.\re{etaF2}, a potentially large fermion anomalous dimension could
strongly modify the UV behavior. Even though this may not happen in
the QED universality class, a fermionic system with large $\ey$ and
strong UV momentum dependences can offer new routes to UV completion
of interacting QFT's. Secondly, it seems worthwhile to extend our
studies to theories with strong NJL-like interactions. The UV flow of
systems with strong gauge and four-fermion couplings still is unknown
territory, the exploration of which is dedicated to future work.

The authors are grateful to Christian Fischer for useful discussions
and acknowledge financial support by the DFG under contract Gi 328/1-2.

\end{document}